\documentclass[runningheads,letter]{llncs}

\usepackage{amssymb}
\usepackage{graphicx,color}

\usepackage{epsfig,graphicx,amsmath,amssymb,lipsum,enumitem,formatfig}
\usepackage{subfigure}
\usepackage{subfig}
\usepackage{algorithm}
\usepackage{algorithmic}

\usepackage{url}

\newcommand{\keywords}[1]{\par\addvspace\baselineskip
  \noindent\keywordname\enspace\ignorespaces#1}

\begin{document}

\mainmatter  

\title {A Mixed-Supervision Multilevel GAN Framework for Image Quality Enhancement}

\titlerunning {Mixed-Supervision Multilevel GAN for Image Quality Enhancement}

\author {Uddeshya Upadhyay \and Suyash P. Awate\thanks{We thank
    support from Aira Matrix and the Infrastructure Facility for
    Advanced Research and Education in Diagnostics grant funded by
    Department of Biotechnology, Government of India
    (RD/0117-DBT0000-002).}}

\index{Upadhyay, Uddeshya}
\index{Awate, Suyash P}

\institute{Computer Science and Engineering, Indian Institute of Tehnology, Bombay, India}



\maketitle

\begin{abstract}

  Deep neural networks for image quality enhancement typically need large quantities of highly-curated training data comprising pairs of low-quality
  images and their corresponding high-quality images.
  While high-quality image acquisition is typically expensive and time-consuming, medium-quality images are faster to acquire, at lower equipment
  costs, and available in larger quantities.
  Thus, we propose a {\em novel generative adversarial network} (GAN) that can leverage {\em training data at multiple levels of quality} (e.g., high
  and medium quality) to improve performance while {\em limiting costs of data curation}.
  We apply our {\em mixed-supervision} GAN to (i)~super-resolve histopathology images and (ii)~enhance laparoscopy images by combining
  super-resolution and surgical smoke removal.
  Results on large clinical and pre-clinical datasets show the benefits of our mixed-supervision GAN over the state of the art.

  
\end{abstract}

\section {Introduction and Related Work}

Image quality enhancement using deep neural networks (DNNs) typically needs large quantities of highly-curated training data comprising corresponding
pairs of low-quality and high-quality images. In this paper, ``low-quality'' images refer to images that have low spatial resolution and exhibit other
degradations; ``high-quality'' images refer to high-resolution uncorrupted images.
While high-quality image acquisition is typically expensive and time-consuming, medium-quality images are faster to acquire, at lower equipment costs,
and available in larger quantities.
Thus, we propose a novel generative adversarial network (GAN) that can leverage {\em training data at multiple levels of quality} (e.g., high and
medium quality) to improve performance while {\em limiting costs of data curation}.

In pre-clinical and clinical digital histopathology, acquisition times increase quadratically with decreasing pixel
width~\cite{fast_acqui,huisman2010creation}. Furthermore, higher-resolution digital scanners are more expensive.
Super-resolution algorithms can enable faster scanning at lower resolutions by filling-in the fine detail in a post-processing step.
{\em Mixed-supervision} can reduce the need for data at the highest resolution and~/ or improve performance using medium-resolution training data; our
earlier works~\cite{Shah2018MICCAI,Shah2019MIDL} propose mixed-supervision for image segmentation.
A class of methods for super-resolution rely on neighbor embeddings~\cite{sr}, sparse representation~\cite{Mousavi2017,Yang2010}, and random
forests~\cite{Schulter2015}.
DNNs have been successful at super-resolution with their ability to optimize features and the regression mapping
jointly~\cite{Dong2014,Bruna2016,Lai2017}.
The class of DNNs giving among the best performances for super-resolution involve GANs~\cite{srgan}.
Our earlier work~\cite{Upadhyay2019} on image super-resolution proposed loss functions for robust learning in the presence of corrupted training data.
Some methods~\cite{Wang2018} use a sequence of GANs for image enhancement (without super-resolution), but, unlike our approach, train each GAN
independently, without defining a unified loss function.
However, none of these methods leverage training data of multiple qualities for (learning) super-resolution.

In {\em laparoscopy}, higher-resolution imaging can offer the surgeon wider views and fine details, both, in the same frame, without needing to move
the endoscope back (to get a larger field of view) and forth (to get the fine details)~\cite{Yamashita2016ultra}. While some state-of-the-art
laparoscopic imaging offers 8K ultra-high-definition (UHD) images, typical and affordable equipment offers lower resolution (less than 2K)
images. Laparoscopic images also suffer from degradation because of surgical smoke.
Super-resolution algorithms can enable low-cost equipment to produce high-resolution images. Smoke-removal algorithms, as well as super-resolution
algorithms, can enhance the performance of subsequent processing involving tracking, segmentation, depth analysis, stereo vision, and augmented
reality, in addition to improving the surgeon's visibility.
Mixed-supervision learning can reduce the need for UHD training sets.
Most methods for image desmoking~\cite{Baid2017,Luo2017,Tchaka2018} model the degradation analytically and propose algorithms to undo the
degradation. On the other hand, a very recent approach~\cite{Chen2018} relies on learning using a DNN.
In contrast, we propose a novel mixed-supervision GAN framework to leverage training data of varying image qualities to improve performance in {\em
  super-resolution coupled with image restoration}.

%
We propose a novel mixed-supervision GAN framework for image quality enhancement that can leverage {\em training data at multiple levels of quality},
e.g., high and medium quality, to improve performance while limiting costs of data curation.
To the best of our knowledge, our framework is the first to propose image quality enhancement using data at multiple quality levels simultaneously.
We apply our framework for (i)~{\em super-resolution in histopathology} images and (ii)~image quality enhancement in {\em laparoscopy images} by
combining {\em super-resolution and surgical smoke removal}.
Results on large clinical and pre-clinical datasets show the benefits of our mixed-supervision GAN over the state of the art.

\section {Methods}

We describe our framework's architecture, loss functions, and the training scheme.

\begin{figure}[!t]
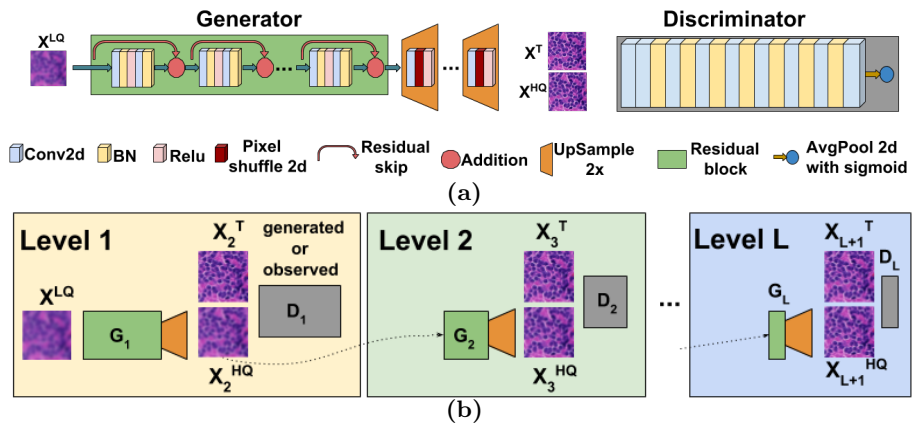

  \oneAcrossTwoColumnsLabela {Pics/MLSRGAN}   {\bf (a)}
  \oneAcrossTwoColumnsLabela {Pics/MLSRGAN_1} {\bf (b)}
  \vspace{-4pt}
  \caption
      {
        {\bf Our Mixed-Supervision Multilevel GAN Framework for Quality Enhancement: MLQEGAN.}
        \textbf{(a)}~Architecture for generator and discriminator, used at each level, in
        \textbf{(b)}~our multilevel architecture with mixed supervision.
      }
      \label{fig:arch}
\end{figure}

{\bf Our Mixed-Supervision Multilevel GAN Architecture.}
While our framework is {\em not} specific to a DNN architecture, we choose to use the generator and the discriminator designs used in
SRGAN~\cite{srgan} (Figure~\ref{fig:arch}(a)). However, our architecture (Figure~\ref{fig:arch}(b)) incorporates a sequence of generators and
discriminators, with each level dealing with increasingly higher quality training data, compared to the low-quality input data.
We refer to a SRGAN-like framework, equivalent to one level in our sequence, as {\em quality-enhancement GAN (QEGAN)}.
We call our framework as {\em multilevel QEGAN (MLQEGAN)} (Figure~\ref{fig:arch}).

For {\em QEGAN}, let the generator $\mathcal{G} (\cdot; \theta)$, parametrized by set $\theta$, take low-resolution and degraded (i.e., low-quality)
input $X^{\text{LQ}}$ and produce super-resolved and restored (i.e., high-quality) output $X^{\text{HQ}} := \mathcal{G} (X^{\text{LQ}}; \theta)$ to
match the observed high-quality true image $X^{\text{T}}$.
Following the GAN learning principle, let the discriminator $\mathcal{D} (\cdot; \phi)$, parametrized by set $\phi$, learn to discriminate between the
probability density function (PDF) of super-resolved and restored generator outputs $P (X^{\text{HQ}})$ and the PDF of observed high-quality true
images $P(X^{\text{T}})$.

For our {\em MLQEGAN}, let there be $L$ levels starting from level $1$ to level $L$.
Let level $l$ have generator $\mathcal{G}_l (\cdot; \theta_l)$ and discriminator $\mathcal{D}_l (\cdot; \phi_l)$.
Let the training data be in the form of random-vector image pairs $(Y^{\text{LQ}}, Y^{\text{T}}_j)$, where
(i)~$Y^{\text{LQ}}$ is the lowest-quality image and
(ii)~$Y^{\text{T}}_j$ is the higher-quality image at quality level $j$ with $j \in [2,L+1]$.
In this paper, the levels in our MLQEGAN framework correspond to different resolutions~/ pixel sizes.
Typical applications lead to training data in large quantities at medium quality as compared to data at higher quality. Thus, our architecture has
progressively fewer parameters to optimize at higher levels, consistent with low sample sizes for higher quality training data.

{\bf Training Set}.
For training MLQEGAN, a higher-quality training image $Y^{\text{T}}_j$ at level $j$ can inform the training of generators and discriminators at all
levels $i < j$.
Thus, from every pair of the form $(Y^{\text{LQ}}, Y^{\text{T}}_j)$, we create {\em multiple} training instances $(X^{\text{LQ}}, X^{\text{T}}_m)$ for
all levels $m \le j$, where
(i)~$X^{\text{LQ}} := Y^{\text{LQ}}$ and
(ii)~$X^{\text{T}}_m$ is the lower-resolution (corresponding to the pixel size at level $m$) version of $Y^{\text{T}}_j$; for $j = m$ we have
$X^{\text{T}}_m = Y^{\text{T}}_m$.
Thus, for training MLQEGAN, we use an {\em effective training set} of the form $\{ (X^{\text{LQ}}, X^{\text{T}}_m) \}$ for all $m \in [2,L+1]$.
Let the set of parameters to be optimized be $\theta := \{ \theta_l \}_{l=1}^L$ and $\phi := \{ \phi_l \}_{l=1}^L$.
Let the union of the generator parameters at levels from $1$ to $l$ be $\theta_1^l := \{ \theta_k \}_{k=1}^l$.

{\bf Loss Functions}.
We design a loss function $\mathcal{L} (\theta, \phi)$ as the sum of loss functions corresponding to each level $l \in [1,L]$.
The loss function at level $l$ comprises
a fidelity loss $\mathcal{L}^F_l (\theta_1^l)$ and
an adversarial loss $\mathcal{L}^A_l (\theta_l, \phi_l)$.
The generator $\mathcal{G}_1 (\cdot; \theta_1)$ at level $1$ takes the lowest-quality input image $X^{\text{LQ}}$ and maps it to $X^{\text{HQ}}_2 :=
\mathcal{G}_1 (X^{\text{LQ}}; \theta_1)$.
Let $\mathcal{G}_1^j := \mathcal{G}_j \circ \mathcal{G}_{j-1} \circ \cdots \circ \mathcal{G}_1$ be the composition of the sequence of generators from
level $1$ to level $j \in [2,L]$.
The generator $\mathcal{G}_l (\cdot; \theta_l)$ at level $l$ takes the lower-quality input image $\mathcal{G}_1^{l-1} (X^{\text{LQ}}; \theta_1^{l-1})$
and maps it to $X^{\text{HQ}}_{l+1} := \mathcal{G}_l (\mathcal{G}_1^{l-1} (X^{\text{LQ}}; \theta_1^{l-1}); \theta_l) = \mathcal{G}_1^l
(X^{\text{LQ}}); \theta_1^l)$.
We propose a {\em fidelity loss} between the generator output $X^{\text{HQ}}_{l+1}$ and the higher-quality image $X^{\text{T}}_{l+1}$ as
\begin{flalign*}
  \mathcal{L}^F_l (\theta_1^l)
  :=
  E_{P (X^{\text{LQ}}, X^{\text{T}}_{l+1})}
  [
    F (\mathcal{G}_1^l (X^{\text{LQ}}; \theta_1^l), X^{\text{T}}_{l+1})
  ]
  ,
\end{flalign*}  
where $F (A,B)$ measures the dissimilarity between images $A$ and $B$; in this paper, $F (A,B)$ is the mean squared error (MSE).
The adversarial loss is the Kullback-Leibler divergence between
(i)~the one-hot probability vectors (distributions) for the generated (``fake'') image $X^{\text{HQ}}_{l+1}$ and true (``real'') image
$X^{\text{T}}_{l+1}$, i.e., $[0,1]$ or $[1,0]$, and
(ii)~the probability vectors (distributions) for the generator-output image $X^{\text{HQ}}_{l+1}$ and the higher-quality image $X^{\text{T}}_{l+1}$,
i.e.,
$[ \mathcal{D}_l (X^{\text{HQ}}_{l+1}; \phi_l), 1 - \mathcal{D}_l (X^{\text{HQ}}_{l+1}; \phi_l) ]$ or
$[ \mathcal{D}_l (X^{\text{T}}_{l+1} ; \phi_l), 1 - \mathcal{D}_l (X^{\text{T}}_{l+1}; \phi_l) ]$,
respectively. We propose the {\em adversarial loss}
\begin{flalign*}
  \mathcal{L}^A_l (\theta_1^l, \phi_l)
  :=
  E_{ P (X^{\text{LQ}}, X^{\text{T}}_{l+1}) }
  [
    \log (1 - \mathcal{D}_l (\mathcal{G}_1^l (X^{\text{LQ}}; \theta_1^l); \phi_l))
    +
    \log (    \mathcal{D}_l (X^{\text{T}}_{l+1} ; \phi_l) )
  ]
  .
\end{flalign*}
We propose the overall loss $L (\theta, \phi) := \sum_{l=1}^L \lambda_l ( \mathcal{L}^F_l (\theta_l^l) + \alpha_l \mathcal{L}^A_l (\theta_1^l,
\phi_l))$, where we fix $\lambda_1 := 1$ and tune the free parameters $\{ \lambda_l \}_{l=2}^L \cup \{ \alpha_l \}_{l=1}^L$ using cross validation.

{\bf Training Scheme}.
We initialize the parameters $\theta \cup \phi$ sequentially, as follows.
We first initialize $\theta_1 \cup \phi_1$ using the training subset with image pairs of the form $(X^{\text{LQ}}, X^{\text{T}}_2)$ to minimize the
loss function $\mathcal{L}^F_1 (\theta_1) + \alpha_1 \mathcal{L}^A_1 (\theta_1, \phi_1)$.
Then, for increasing levels $l$ from $2$ to $L$, we initialize $\theta_l \cup \phi_l$ using the training subset $(X^{\text{LQ}}, X^{\text{T}}_l)$ to
minimize the loss function $\mathcal{L}^F_l (\theta_l^l) + \alpha_l \mathcal{L}^A_l (\theta_1^l, \phi_l)$, but fixing all previous-level generator
parameters $\theta_1^{l-1}$.
After initialization, we train all GANs in a joint optimization framework using Adam~\cite{Adam}, with internal parameters $\beta_1 := 0.9, \beta_2 :=
0.999$, initial learning rate $0.002$ that decays based on cosine annealing, batch size $8$, and number of epochs $500$.

\section {Results and Discussion}
\label {sec:results}

We evaluate our mixed-supervision MLQEGAN to (i)~super-resolve histopathology images and (ii)~enhance laparoscopy images by combining super-resolution
and surgical smoke removal.
We compare against a SRGAN-like architecture that is the among the state of the art for super-resolution, which we leverage additionally for quality
enhancement, i.e., QEGAN.
To quantitate performance, we use three complementary measures:
(i)~relative root MSE (RRMSE) between the DNN output, say, $A^{\text{HQ}}$ and the corresponding true image, say, $A^{\text{T}}$, as $\| A^{\text{HQ}}
- A^{\text{T}} \|_2 / \| A^{\text{T}} \|_2$,
(ii)~multiscale structural similarity index (msSSIM)~\cite{ssim}, and
(iii)~quality index based on local variance (QILV)~\cite{QILV}.
While msSSIM is more sensitive to image noise than blur, QILV is more sensitive to image blur instead.
In this paper, the cross-validation tuned free-parameter values are: $\forall l, \alpha_l = 3e^{-5}$;
when $L$$=$$2$, $\lambda_1 = 1e^{-4}$, $\lambda_2 = 1$;
when $L$$=$$3$, $\lambda_1 = 1e^{-4}$, $\lambda_2 = 1^{e-4}$, $\lambda_3 = 1$.

\begin{figure}[!t]
  \fourAcrossLabelsHeight {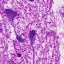} {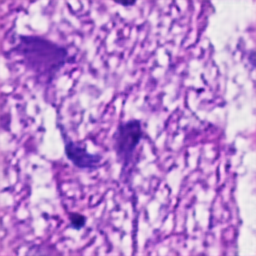} {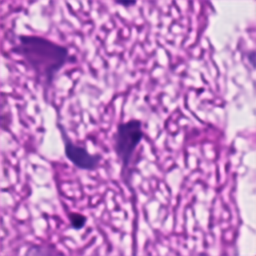} {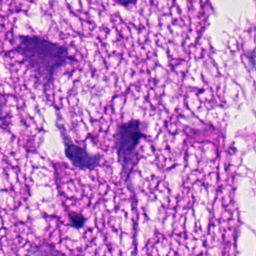} {{\bf (a1)} Low Res. $64^2$} {{\bf (b1)} QEGAN} {{\bf (c1)} Our MLQEGAN} {{\bf (d1)} Truth $256^2$} {1.03}
  \fourAcrossLabelsHeight {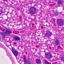} {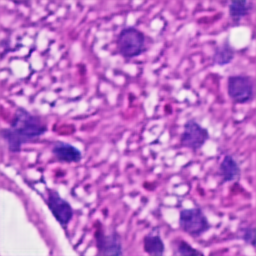} {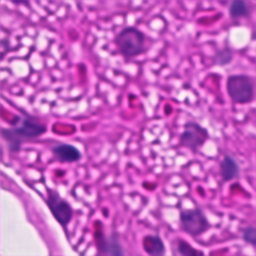} {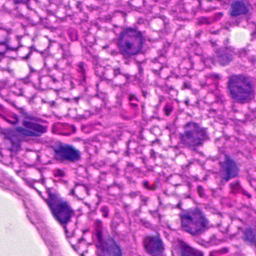} {{\bf (a2)} Low Res. $64^2$} {{\bf (b2)} QEGAN} {{\bf (c2)} Our MLQEGAN} {{\bf (d2)} Truth $256^2$} {1.03}
  \vspace{-4pt}
  \caption
      {
        \textbf{Results on Histopathology Images.}
        \textbf{(a1)-(a2)}~Low-resolution 64$^2$ (from 10$\times$ magnification).
        (RRMSE, msSSIM, QILV) for:
        \textbf{(b1)-(b2)}~QEGAN 4$\times$ super-resolution (256$^2$):             (0.131, 0.868, 0.830), (0.117, 0.837, 0.919).
        \textbf{(c1)-(c2)}~{\bf Our MLQEGAN} 4$\times$ super-resolution (256$^2$): (0.117, 0.910, 0.90), (0.091, 0.922, 0.942).
        \textbf{(d1)-(d2)}~Truth 256$^2$ (from 40$\times$ magnification).
      }
      \label {fig:resultsHistopathology}
\end{figure}

\begin{figure}[!t]
  \threeAcrossHeightLabels {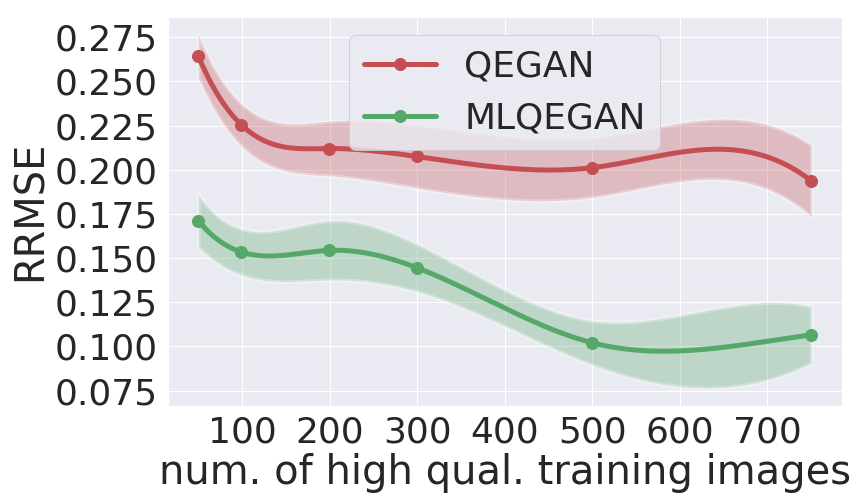} {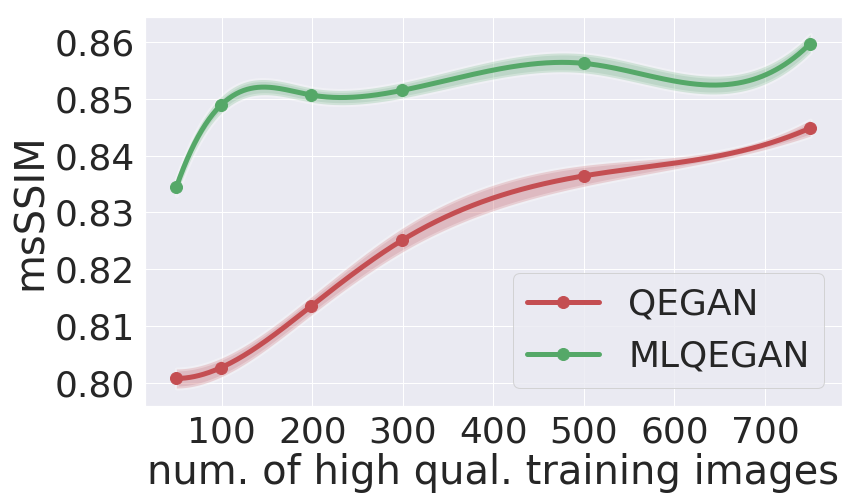} {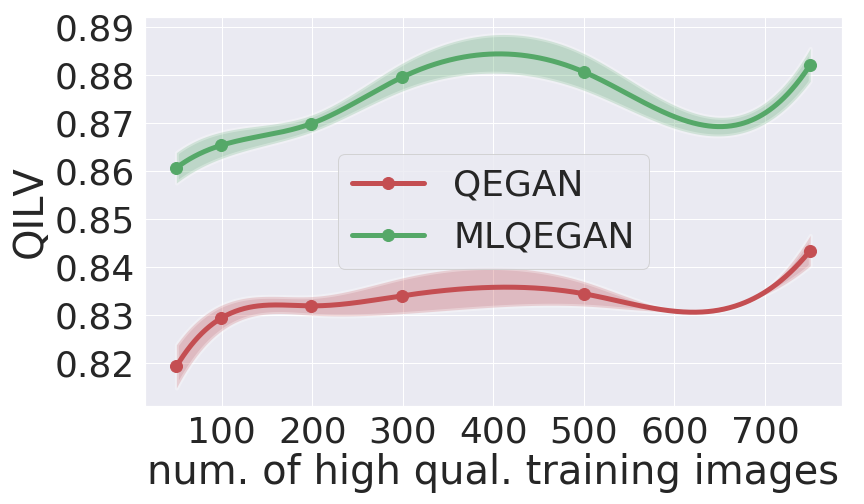} {{\bf (a1)} Histopath. 4$\times$SR} {{\bf (a2)} Histopath. 4$\times$SR} {{\bf (a3)} Histopath. 4$\times$SR} {0.204}
  \threeAcrossHeightLabels {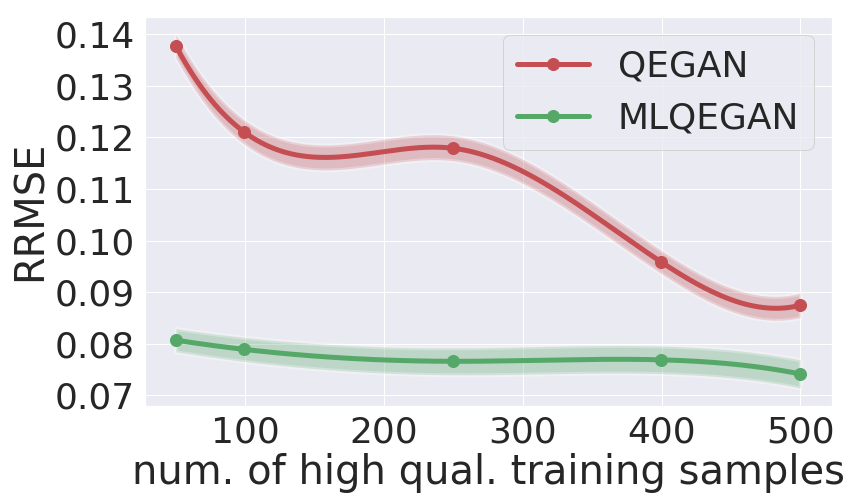} {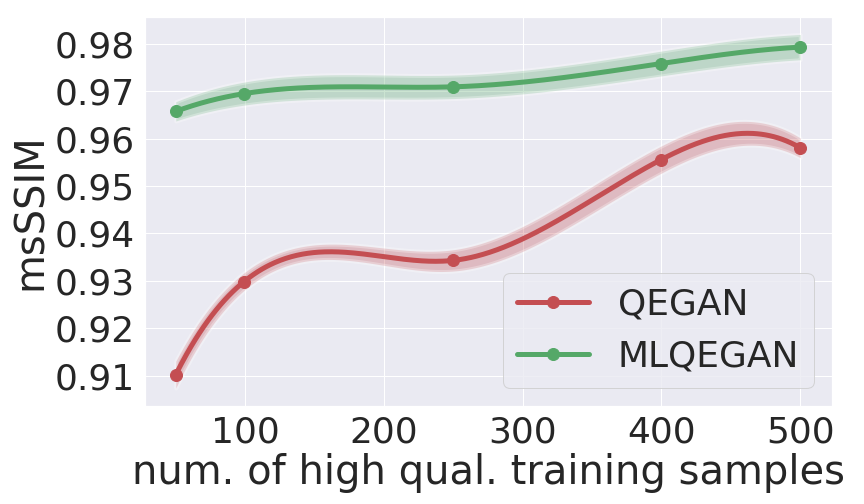} {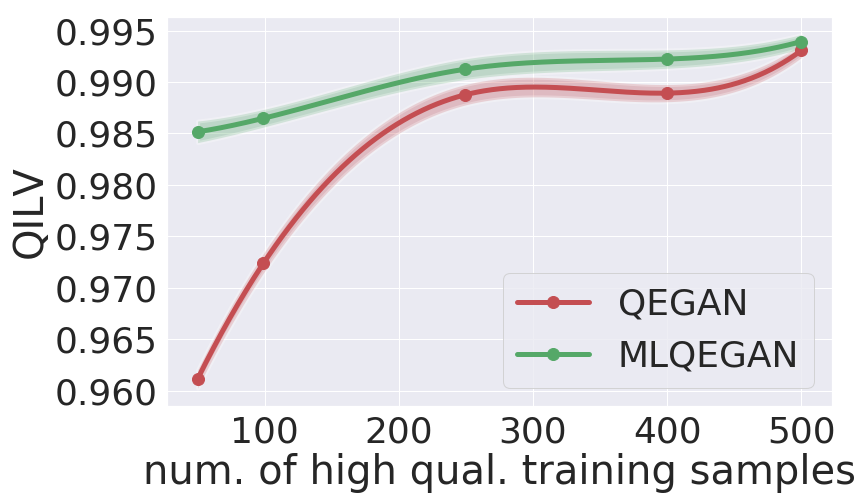} {{\bf (b1)} Lap. 4$\times$SR+Desmoke} {{\bf (b2)} Lap. 4$\times$SR+Desmoke} {{\bf (b3)} Lap. 4$\times$SR+Desmoke} {0.204}
  \threeAcrossHeightLabels {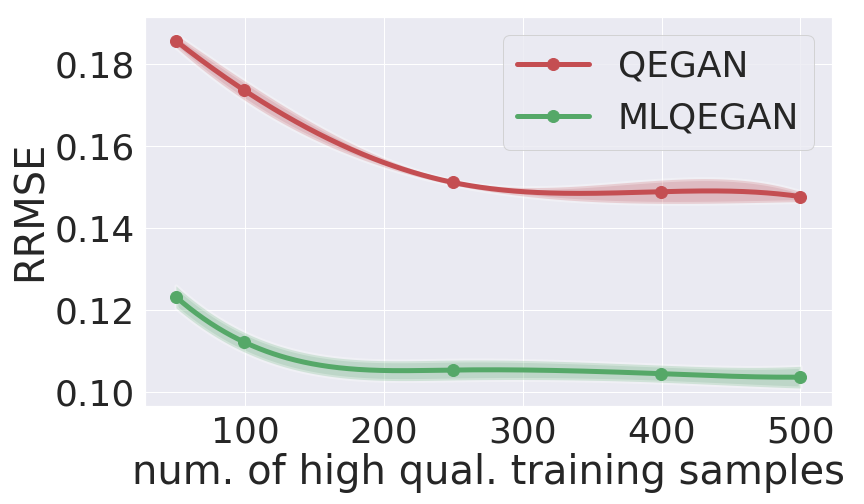} {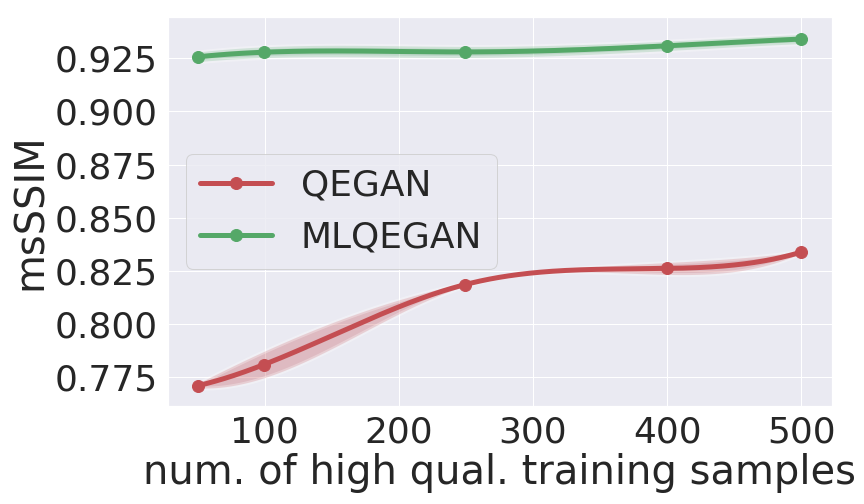} {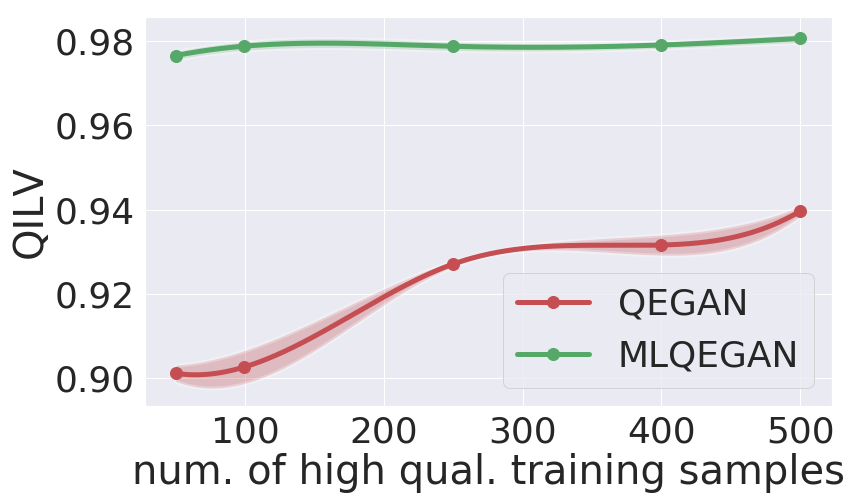} {{\bf (c1)} Lap. 8$\times$SR+Desmoke} {{\bf (c2)} Lap. 8$\times$SR+Desmoke} {{\bf (c3)} Lap. 8$\times$SR+Desmoke} {0.204}
  \vspace{-4pt}
  \caption
      {
        \textbf {Results on Histopathology and Laparoscopy Images: Quantitative Analysis.}
        RRMSE, msSSIM, QILV for images in:
        \textbf{(a1)-(a3)} histopathology: 4$\times$ super-resolution;
        \textbf{(b1)-(b3)} laparoscopy: 4$\times$ super-resolution + desmoking;
        \textbf{(c1)-(c3)} laparoscopy: 8$\times$ super-resolution + desmoking.
        The error bars indicate the variability from randomness in data sampling and Adam~\cite{Adam}.
      }
      \label {fig:resultsQuant}
\end{figure}

\textbf {Results on Histopathology Images.}
We take $8$ whole-slide histopathology images from pre-clinical Wistar rat biopsies at 40$\times$ magnification from a Hamamatsu scanner, which
includes images at multiple magnification levels, including 10$\times$ and 20$\times$.
The task is to map the low-quality input data $X^{\text{LQ}}$ at 10$\times$ to the high-quality data $X^{\text{T}}$ at 40$\times$.
MLQEGAN uses $L$$=$$2$ levels.

The {\em training set} comprises image-pair instances of $(Y^{\text{LQ}}, Y^{\text{T}}_2)$ and $(Y^{\text{LQ}}, Y^{\text{T}}_3)$.
We create instances of $(Y^{\text{LQ}}, Y^{\text{T}}_2)$ by
(i)~randomly selecting $5000$ non-overlapping patches at 20$\times$ (128$^2$ pixels) to give $Y^{\text{T}}_2$;
(ii)~selecting the corresponding patches at 10$\times$ (64$^2$ pixels) to give $Y^{\text{LQ}}$.
We create instances of $(Y^{\text{LQ}}, Y^{\text{T}}_3)$ by
(i)~randomly selecting non-overlapping patches at 40$\times$ (256$^2$ pixels) to give $Y^{\text{T}}_3$;
(ii)~selecting the corresponding patches at 10$\times$ (64$^2$ pixels) to give $Y^{\text{LQ}}$.
We vary the number of the highest-quality image-pair instances of $(Y^{\text{LQ}}, Y^{\text{T}}_3)$ from $50$ to $750$.

We create the {\em validation set} of image-pair instances of $(V^{\text{LQ}}, V^{\text{T}}_3)$ by randomly choosing $100$ images at 40$\times$ to
give $V^{\text{T}}_3$, and their corresponding patches at 10$\times$ to give $V^{\text{LQ}}$.
We similarly create the {\em test set} of $1000$ instances of $(Z^{\text{LQ}}, Z^{\text{T}}_3)$.

While QEGAN can leverage only a small subset of available training data comprising instances of the form $(Y^{\text{LQ}}, Y^{\text{T}}_3)$, our
MLQEGAN is able leverage the entire training set including instances of the forms $(Y^{\text{LQ}}, Y^{\text{T}}_3)$ and $(Y^{\text{LQ}},
Y^{\text{T}}_2)$.
Our MLQEGAN outputs (Figure~\ref{fig:resultsHistopathology}(c1)-(c2)) are much closer to the ground truth
(Figure~\ref{fig:resultsHistopathology}(d1)-(d2)), compared to QEGAN outputs (Figure~\ref{fig:resultsHistopathology}(b1)-(b2)).  Unlike QEGAN, MLQEGAN
is able to extract useful information in the medium-quality images $Y^{\text{T}}_2$ available in significantly larger quantities, compared to the
highest-quality images $Y^{\text{T}}_3$. In this way, MLQEGAN clearly outperforms QEGAN in reproducing the true textural appearances (colors and
features) and cell shapes.
Quantitatively (Figure~\ref{fig:resultsQuant}(a1)--(a3)), while QEGAN's performance reduces steadily with the reduction in the number of training-set
images at the highest-quality level, our MLQEGAN's performance remains much more stable. Moreover, our MLQEGAN's performance consistently stays
significantly better than QEGAN's performance, for all training-set sizes.

\begin{figure}[!t]
  \fourAcrossLabelsHeight {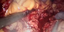} {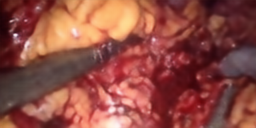} {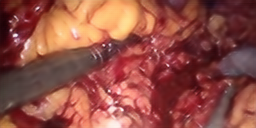} {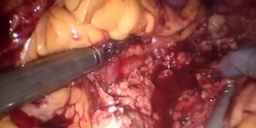} {{\bf (a1)}Low Qual. 32$\times$64} {{\bf (b1)}QEGAN} {{\bf (c1)}Our MLQEGAN} {{\bf (d1)}Truth 128$\times$256} {0.52}
  \fourAcrossLabelsHeight {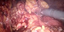} {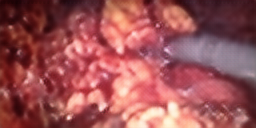} {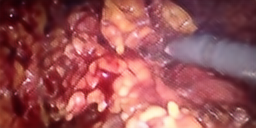} {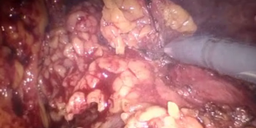} {{\bf (a2)}Low Qual. 32$\times$64} {{\bf (b2)}QEGAN} {{\bf (c2)}Our MLQEGAN} {{\bf (d2)}Real 128$\times$256} {0.52}
  \fourAcrossLabelsHeight {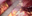} {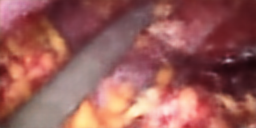} {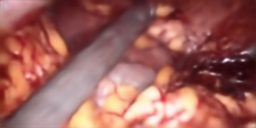} {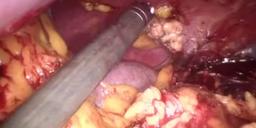} {{\bf (a3)}Low Qual. 16$\times$32} {{\bf (b3)}QEGAN} {{\bf (c3)}Our MLQEGAN} {{\bf (d3)}Truth 128$\times$256} {0.52}
  \fourAcrossLabelsHeight {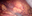} {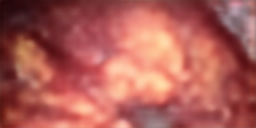} {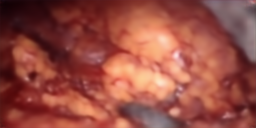} {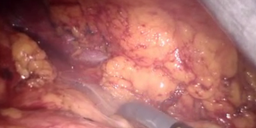} {{\bf (a4)}Low Qual. 16$\times$32} {{\bf (b4)}QEGAN} {{\bf (c4)}Our MLQEGAN} {{\bf (d4)}Real 128$\times$256} {0.52}
  \vspace{-4pt}
  \caption
      {
        {\bf Results on Laparoscopic Images: Super-Resolution and Desmoking}.
        (RRMSE, msSSIM, QILV) are in parentheses.
        {\bf (a1)}~{\em Input with 4$\times$ lower-resolution, simulated smoke};
        {\bf (b1)}~QEGAN: (0.112,0.897,0.963);
        {\bf (c1)}~{\bf Our MLQEGAN}: (0.065,0.971,0.993);
        {\bf (d1)}~Truth.
        {\bf (a2)}~{\em Input with 4$\times$ lower-resolution, real smoke};
        {\bf (b2)}~QEGAN;
        {\bf (c2)}~{\bf Our MLQEGAN};
        {\bf (d2)}~Real smoky high-resolution image.
        {\bf (a3)}~{\em Input with 8$\times$ lower-resolution, simulated smoke};
        {\bf (b3)}~QEGAN: (0.151,0.825,0.926);
        {\bf (c3)}~{\bf Our MLQEGAN}: (0.098,0.925,0.988);
        {\bf (d3)}~Truth.
        {\bf (a4)}~{\em Input with 8$\times$ lower-resolution, real smoke};
        {\bf (b4)}~QEGAN;
        {\bf (c4)}~{\bf Our MLQEGAN};
        {\bf (d4)}~Real smoky high-resolution image.
      }
      \label {fig:resultsLaparoscopy}
\end{figure}

{\bf Results on Laparoscopic Images}.
We use the da Vinci dataset~\cite{daVinci}.
Our task involves smoke removal and 4$\times$ super-resolution.
MLQEGAN uses $L$$=$$2$.

The {\em training set} comprises image-pair instances of $(Y^{\text{LQ}}, Y^{\text{T}}_2)$ and $(Y^{\text{LQ}}, Y^{\text{T}}_3)$.
We create instances of $(Y^{\text{LQ}}, Y^{\text{T}}_2)$ by
(i)~randomly selecting $5000$ smokeless full-resolution images and 2$\times$ downsampling them to give $Y^{\text{T}}_2)$;
(ii)~degrading the same selected image set with smoke, and then 4$\times$ downsampling (we smooth a bit using Gaussian convolution before downsampling
to prevent aliasing) to create the low-quality training set $Y^{\text{LQ}}$.
We create instances of $(Y^{\text{LQ}}, Y^{\text{T}}_3)$ by
(i)~randomly selecting full-resolution smokeless images to give $Y^{\text{T}}_3$;
(ii)~degrading and downsampling the same selected image set to give $Y^{\text{LQ}}$.
We vary the number of the highest-quality image pairs $(Y^{\text{LQ}}, Y^{\text{T}}_3)$ from $50$ to $500$.

We create the {\em validation set} of image-pair instances of $(V^{\text{LQ}}, V^{\text{T}}_3)$ by randomly choosing $100$ images $V^{\text{T}}_3$ and
creating $V^{\text{LQ}}$ by degrading and downsampling as before.
We similarly create the {\em test set} of $1000$ instances of $(Z^{\text{LQ}}, Z^{\text{T}}_3)$.

While QEGAN learning is unable to use medium-quality images, our MLQEGAN uses them to improve performance significantly.
With simulated smoke, results from our MLQEGAN (Figure~\ref{fig:resultsLaparoscopy}(c1)) appear sharper than those from QEGAN
(Figure~\ref{fig:resultsLaparoscopy}(b1)).
We also evaluate the trained models for QEGAN and MLQEGAN on real-world smoke by selecting {\em smoky frames}
(Figure~\ref{fig:resultsLaparoscopy}(d2)) and downsampling them by 4$\times$ to create the low-quality input
(Figure~\ref{fig:resultsLaparoscopy}(a2)).
The results from MLQEGAN (Figure~\ref{fig:resultsLaparoscopy}(c2)) show sharper and more realistic textures.
Analogous to the aforementioned experiments, we tested all methods in a very challenging scenario with {\em 8$\times$ downsampling}, where MLQEGAN
uses $L$$=$$3$ levels. Here as well, MLQEGAN performs better than QEGAN (Figure~\ref{fig:resultsLaparoscopy}(a3)-(d4)).

Quantitatively, with simulated smoke, for 4$\times$ and 8$\times$ downsampling (Figure~\ref{fig:resultsQuant}(b1)-(c3)) shows that MLQEGAN outperforms
QEGAN in all measures.

{\bf Time Requirements.}
Our MLQEGAN has fewer layers and fewer parameters (by roughly 20\%) than SRGAN (or QEGAN). Training time can depend on many factors, e.g., network
architecture, training set sizes, and epochs needed to converge. Our MLQEGAN uses more training images at lower quality, but with smaller sizes. In
general, our training time is a bit larger than that of QEGAN. For test data, our MLQEGAN produces the output a bit faster than SRGAN (or QEGAN)
because we have fewer parameters and layers.

{\bf Conclusion.}
We proposed a novel {\em mixed-supervision} GAN that leverages training data at multiple levels of quality to improve performance while limiting costs
of data curation.
We propose a novel {\em multilevel} architecture with a sequence of GANs with (i)~progressively decreasing complexity and (ii)~loss functions using
coupled generator sequences.
We apply our mixed-supervision GAN to (i)~{\em super-resolve histopathology} images and (ii)~enhance {\em laparoscopy} images by combining {\em
  super-resolution and surgical smoke removal}.
Results on large datasets show the benefits of our mixed-supervision GAN over the state of the art.

\bibliographystyle{splncs04}
\bibliography{./Bibtex_DNN}

\end{document}